\documentclass[preprint,12pt,groupaddress]{elsarticle}

\usepackage{booktabs}
\usepackage{color}
\usepackage{lipsum}
\usepackage{float}
\usepackage[titletoc,toc,title]{appendix}
\usepackage{amssymb}
\usepackage{epsfig}
\usepackage{amsmath}
\usepackage{times}
\usepackage{subfigure}
\usepackage{graphicx}
\usepackage{booktabs}
\usepackage{colortbl}
\usepackage{natbib}

\begin{document}

\title{On the fractional cell kill law governing the lysis of solid tumors}

\author{\'{A}lvaro G. L\'{o}pez}
\author{Jes\'{u}s M. Seoane}
\author{Miguel A. F. Sanju\'an}

\address{Nonlinear Dynamics, Chaos and Complex Systems Group.\\Departamento de F\'isica, Universidad Rey Juan Carlos, Tulip\'an s/n, 28933 M\'ostoles, Madrid, Spain}

\date{\today}

\begin{abstract}

The \emph{fractional cell kill} is a mathematical expression describing the rate at which a certain population of cells is reduced to a fraction of itself. In order to investigate the fractional cell kill that governs the rate at which a solid tumor is lysed by a cell population of cytotoxic CD8$^{+}$ T cells (CTLs), we present several \emph{in silico} simulations and mathematical analyses. When the CTLs eradicate efficiently the tumor cells, the models predict a correlation between the morphology of the tumors and the rate at which they are lysed. However, when the effectiveness of the immune cells is decreased, the mathematical function fails to reproduce the process of lysis. This limit is thoroughly discussed and a new fractional cell kill is proposed.\\
\end{abstract}

\maketitle



\section{Introduction}\label{sec:intro}

Immunotherapy has been focusing great attention among cancer therapies in the recent past years. Adoptive cell transfer using chimeric antigen receptors \citep{adoptive,chimeric}, the modulation of CTLA-4 activity by means of monoclonal antibodies \citep{ctla4}, or the blocking of the PD-1 receptor \cite{antip1d}, are a few outstanding examples. The progress of tumor immunotherapy with T lymphocytes mainly relies on our capacity to uncover and understand the molecular and cellular basis of the T-cell-mediated antitumor response. However, due to the highly complex regulatory mechanisms that control both cell growth and the immune system, this task can be hardly achieved without the use of mathematical models. From a theoretical point of view, these models provide an analytical framework in which fundamental questions concerning cancer dynamics can be addressed in a rigorous fashion. The practical reason for their development is to make quantitative predictions that permit the refinement of the existing therapies, or even the design of new ones.

Mathematical models of tumor growth and its interaction with the immune system have demonstrated their potential to explain different properties of tumor-immune interactions \citep{momatim}. Among these models, a continuous ODE model was engineered in \cite{immutu} to explore a possible dynamical origin of the dormancy and the sneaking-through of tumors. Such model consists of two cell populations (tumor and immune effector cells), and describes their interaction as an enzymatic process. In particular, these authors consider that the rate at which the tumor is lysed by the cytotoxic cells, \emph{i. e.}, the fractional cell kill, increases linearly with the number of immune cells. In other words, it is possible to increase without bounds the speed at which the tumor is destroyed, by simply adding more immune cells. Following this work, a more specialized model was designed years later \citep{validpilis}, which includes the adaptive (CD8$^{+}$ lymphocytes) and the innate (NK cells) cell-mediated immune responses. This model was validated using experiments in mice \citep{nkg2d} and humans \citep{humans}. In order to reproduce the experimental data, the authors incorporated a new fractional cell kill. More specifically, these authors noticed that the fraction of lysed tumor cells after a few hours, plotted versus different values of the ratio between the initial number of immune and tumor cells, saturates for increasing values of the latter. Consequently, they proposed a sigmoid function \citep{validpilis} depending on the effector-to-target ratio as the mathematical function describing the rate at which a tumor is lysed.

In \cite{validbul} we have developed a simpler model, validated it, and proposed several hypotheses to explain the nature of the fractional cell kill. In particular, it was suggested that the saturation might arise as a consequence of T cell crowding, which depends on the geometry of the tumor. The main purpose of this work is to characterize more rigorously the nature of the mathematical expression that governs the lysis of tumor cells by cytotoxic cells. Our study supports the previous hypothesis, indicating that this mathematical function emerges from spatial and geometrical restraints. Interestingly, simulations are provided in the limit of immunodeficient environments, where saturation becomes less evident. We demonstrate that the current mathematical function works bad for such environments, and retake the conceptual framework of enzyme kinetics to propose another fractional cell kill. We show that this new function behaves better in the limit in which the immune cell population is small compared to the tumor size, and that the parameters appearing in it have a clear physical and biological interpretation.
\begin{figure}
\centering
\begin{tabular}{cc}
\subfigure[]{
  \includegraphics[width=0.48\linewidth,height=0.51\linewidth] {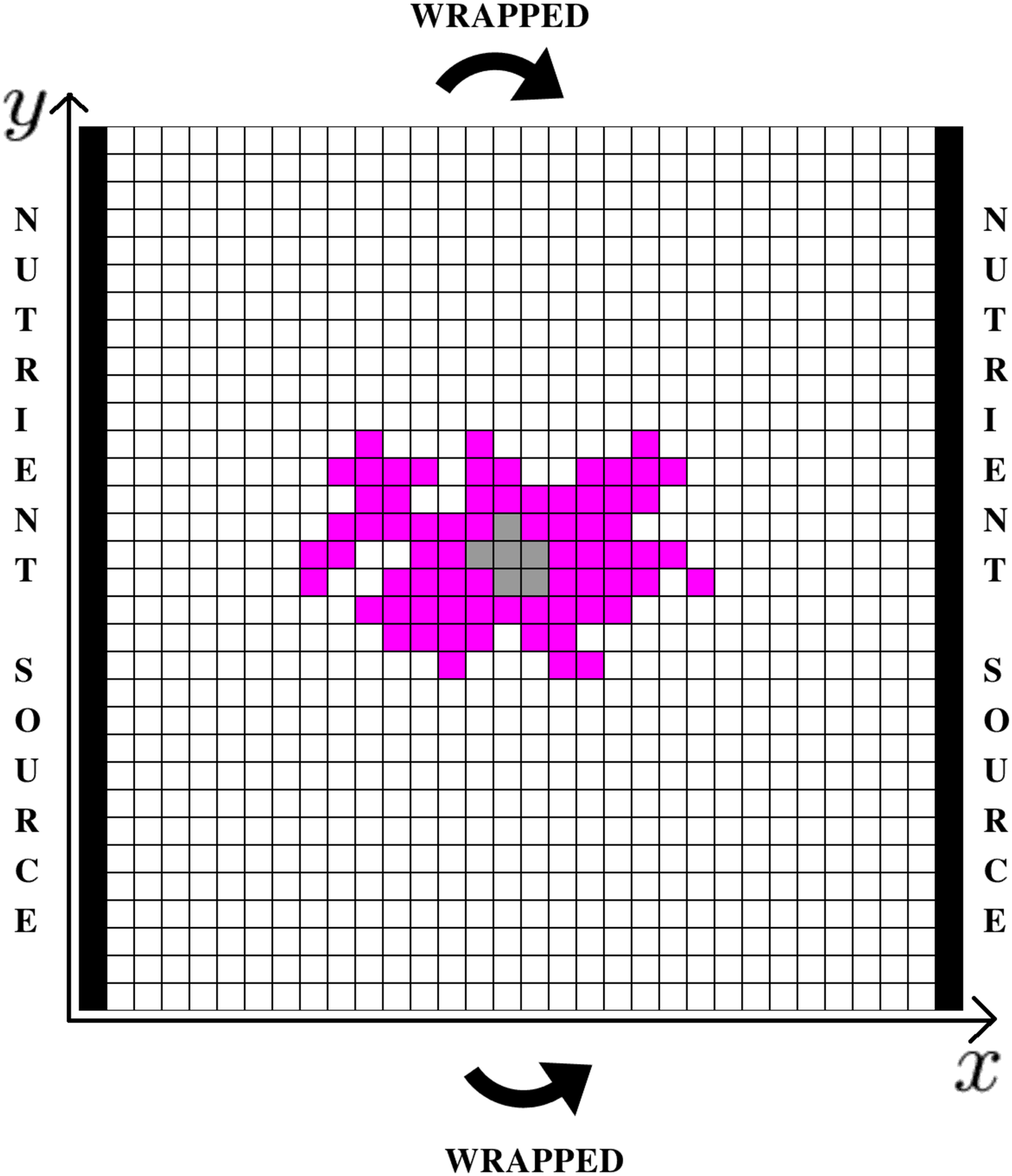}
   \label{subfig:1}
   } &
\subfigure[]{
   \includegraphics[width=0.48\linewidth,height=0.51\linewidth] {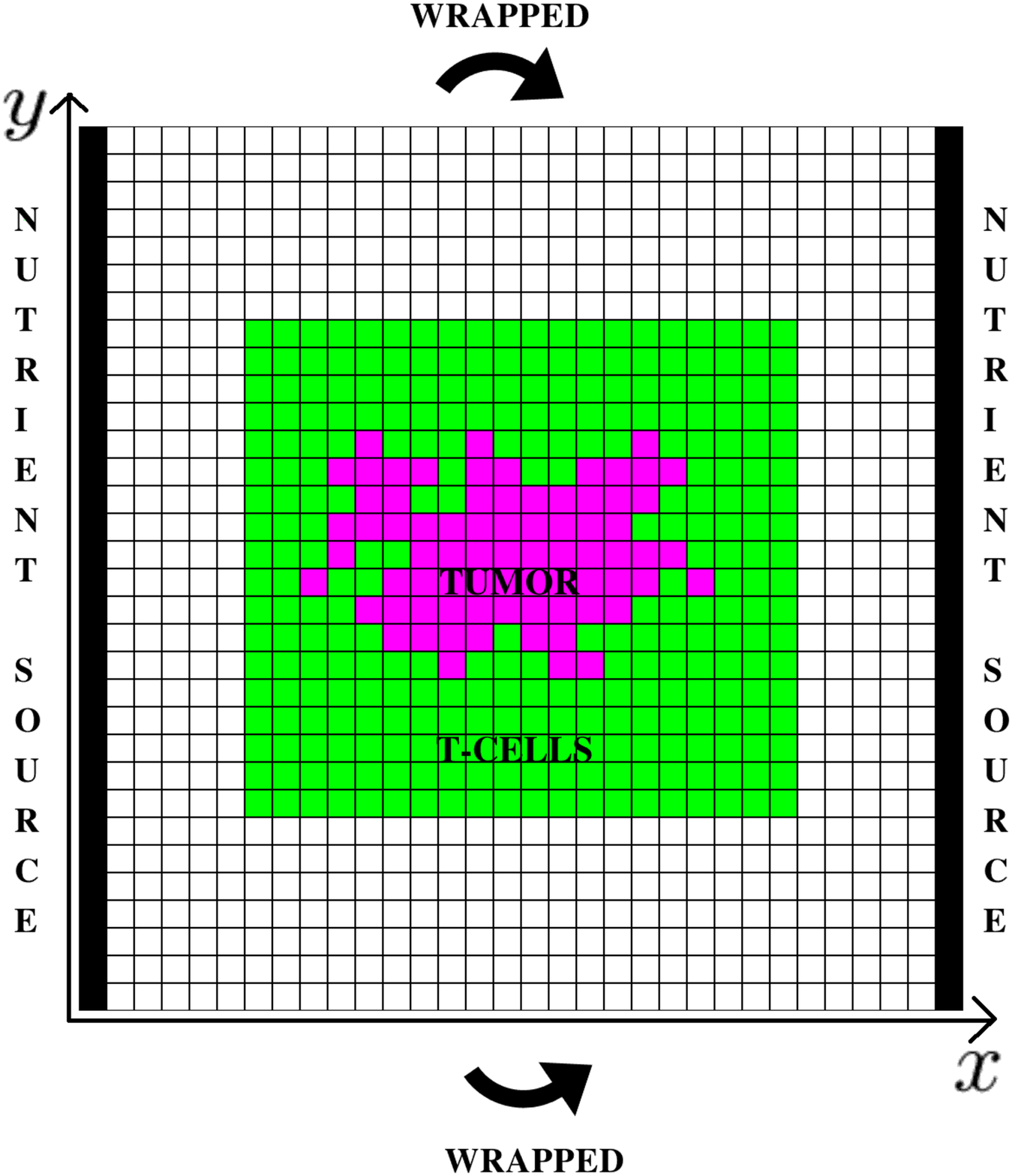}
   \label{subfig:2}
   } 
\end{tabular}
\caption{(a) Schematic representation of the cellular automaton grid in a square domain, with some tumor cells (pink) growing from its center, and some necrotic cells (gray) at its core. Two vertical vessels on the boundary supply the nutrients required for cell division and other cellular activities. The upper and lower bounds are identified, forming a cylinder. (b) To study the lysis of the tumors, the initial conditions are always prepared by randomly placing the effector cells in a rectangular region outside the tumor. The size of this domain is selected so that for the maximum values of the effector-to-target ratio the region is almost filled with effector cells.}
\label{fig:cabas}
\end{figure}

\section{Models}\label{sec:md}
 
\subsection{An hybrid cellular automaton model}

The simulations are accomplished by means of a cellular automaton (CA) model developed in \cite{cahyb} to study the interactions between tumor and immune effector cells. This model was built on a previously CA model designed to study the effects of competition for nutrients and growth factors in avascular tumors \citep{ferreira}. It is hybrid because the cells are treated discretely, allowing them to occupy several grid points in a particular spatial domain, and evolve according to probabilistic and direct rules. On the other hand, the diffusion of nutrients (such as glucose, oxygen and other types of nutrients) from the vessels into such spatial region is represented through linear reaction-diffusion equations, which are continuous and deterministic. Two types of nutrients are utilized in this model, making a distinction between those which are specific for cell division $N(x,y,t)$, and others $M(x,y,t)$ that are related to the remaining cellular activities. The partial differential equations for the diffusion of nutrients are
\begin{eqnarray}
&\dfrac{\partial N}{\partial t}&=D_{N}\nabla^{2}N-k_{1}T N-k_{2}H N-k_{3}E N \label{eq:1} \\ 
&\dfrac{\partial M}{\partial t}&=D_{M}\nabla^{2}M-k_{4}T M-k_{5}H M-k_{6}E M,\label{eq:2} 
\end{eqnarray}
where $T(x,y,t)$, $H(x,y,t)$ and $E(x,y,t)$ are functions representing the number of tumor, healthy and immune cells at time $t$ and position $(x,y)$. For simplicity, we assume that both type of nutrients have the same diffusion coefficient $D_{N}=D_{M}=D$. Following \cite{cahyb}, we consider that the competition parameters are equal $k_{2}=k_{3}=k_{5}=k_{6}=k$, except for the tumor cells, which compete more aggressively. We set $k_{1}=\lambda_{N}k$ and $k_{4}=\lambda_{M}k$, with $\lambda_{M}$ and $\lambda_{N}$ greater than one. An adiabatic limit is considered, assuming that the solutions are stationary. This approximation holds because the time it takes a tumor cell to complete its cell cycle, which is of the order of days \citep{amolap}, is much longer than that of the diffusion of nutrients. A quadrilateral domain $\Omega=[0,L]\times[0,L]$ is considered and Dirichlet boundary conditions are imposed on the vertical sides of the domain, where the vessels are placed, assigning $N(0,y)=N(L,y)=N_{0}$ and $=M(0,y)=M(L,y)=M_{0}$. For simplicity, the horizontal upper and lower bounds of the domain obey periodic boundary conditions $N(x,0)=N(x,L)$ and $M(x,0)=M(x,L)$, wrapping them together to form a cylinder.

Finally, the diffusion equations are nondimensionalized as explained in \cite{ferreira}, and the equations are numerically solved by using finite-difference methods with successive over-relaxation. The resolution of the grid $n$ equals 300 pixels in all our simulations. For a detailed description of the CA algorithm we refere the reader to \cite{ferreira}. The simulations here presented are carried out in two successive steps. The first is devoted to the growth of the tumors, while the second focuses on the lysis of tumor cells by CTLs.
\begin{enumerate}
\item We generate distinct solid tumors as monoclonal growths, arising after many iterations of the cellular automaton. At each CA iteration the tumor cells can divide, move or die attending to certain probabilistic rules that depend on the nutrient concentration per tumor cell and some specific parameters. Each of these parameters $\theta_{a}$ represent the intrinsic capacity of the tumor cells to carry out a particular action $a$. The precise probabilistic laws and the corresponding actions are described in detail in Appendix~\ref{app:caraa}. Attending to morphology, diverse types of tumors can be generated, depending on the nutrient competition parameters among tumor cells $\alpha,\lambda_{N}$. We simulate four types of geometries (spherical, papillary, filamentary and disconnected), and inspect four tumors of different sizes for each shape. 
\item The lysis of tumor cells is a hand-to-hand struggle comprising several processes. After recognition of these cells through antigen presentation via MHC class I molecules, the CD8$^{+}$ T cells proceed to induce apoptosis. The principal mechanism involves the injection of proteases through pores on the cell membrane, that have been previously opened by polymerization of perforins. Even though death may take about an hour to become evident, it takes minutes for a T cell to program antigen-specific target cells to die \citep{bioimm}. We assign a time of ten minutes for each iteration of the CA, and other choices can be made. Therefore, twenty four iterations of the CA equal the four hours after which the lysis of tumor cells is measured in the experiments \citep{nkg2d}. Since the cell cycle time of a tumor cell is generally a few times longer, we assume a second adiabatic approximation and suspend the tumor cell dynamics during T cell lysis. 

The rules governing the effector cells evolution are as follows. At each iteration, those immune cells that are in contact with at least one tumor cell, might lyse them with certain probability. The intrinsic cytotoxic capability, which in the model also accounts for the capacity of T cells to recognize tumor cells \citep{tils}, is related to the parameter $\theta_{lys}$. If a T cell destroys a tumor cell, recruitment might be induced in its neighbouring CA elements. When immune cells are not in direct contact with a tumor cell, they can either move or become inactivated. Thus, the present CA model does not represent T cell infiltration into the tumor mass, which is discussed somewhere else \cite{dedytu}. We consider that a single T cell can not lyse more than three times, leaving the region of interest when this occurs \citep{cahyb}. The precise probabilistic laws and the corresponding actions are again thoroughly described in Appendix~\ref{app:caraa}. Each of the sixteen solid tumors is co-cultivated with different effector-to-target ratios as initial conditions (see Fig.~\ref{fig:cabas}) and the lysis is computed four hours later.
\end{enumerate}

Because our study mainly focuses on how fast lymphocytes lyse a tumor, an important simplification between our cellular automaton and the one presented in \cite{cahyb} deserves notification. We have excluded a constant source of NK cells from the model.

\subsection{An ordinary differential equation model}

In the present investigation the results of the simulations performed with the cellular automaton model are fitted by means of a least-squares fitting method to a Lotka-Volterra type model. The continuous model of cell-mediated immune response to tumor growth consists of three interacting cell populations: the tumor cells $T(t)$, the host healthy cells $H(t)$ and the immune effector cells $E(t)$. Our study focus mainly on CD$8^{+}$ T lymphocytes, but the model can be easily modified to reproduce NK cell dynamics. The system of differential equations \citep{validbul} reads
\begin{eqnarray}
&\dfrac{d T}{d t}&=r_{1}T\left(1-\dfrac{T}{K_{1}}\right)-a_{1 2}H T-K(E,T)T \label{eq:3}\bigskip\\ 
& \dfrac{d H}{d t}&=r_{2}H\left(1-\dfrac{H}{K_{2}}\right)-a_{2 1}T H \label{eq:4}\bigskip\\
& \dfrac{d E}{d t}&=\sigma-d_{3} E+g\dfrac{K^{2}(E,T)T^{2}}{h+K^{2}(E,T)T^{2}}E-a_{3 1} T E\label{eq:5},
\end{eqnarray}
with 
\begin{equation}
K(E,T)=d \dfrac{(E/T)^{\lambda}}{s+ (E/T)^{\lambda}}.
\label{eq:6}
\end{equation}

The tumor cells and the host healthy cells grow logistically with growth rates and carrying capacities $r_{1}, K_{1}$ and $r_{2}, K_{2}$, respectively. The terms $a_{12}$ and $a_{21}$ model the competition for nutrients and space among tumor and healthy cells. A term representing the fractional cell kill of tumor cells by CTLs is given by the nonlinear function $K(E,T)$, which constitutes the main topic of the present work. Here the parameter $\sigma$ incorporates a constant input of lymphocytes into the tissue where the tumor develops, but it can be related to a background of NK cells as well \citep{validpilis}. The inactivation of the effector cells and their migration from the tumor area is given by the term $d_{3}E$, whereas the parameters $g,h$ stand for the recruitment of immune cells to the tumor domain mediated by cytokines, such us IFN-$\gamma$ or TNF-$\alpha$, after the tumor and the immune cells interact. Finally, the competition between the tumor and the T cells for resources is given by $a_{31}$. These differential equations are solved using a fourth order Runge-Kutta integrator.

This continuous model has been validated \citep{validbul} using experiments from \cite{nkg2d} and the parameter values are listed in Table~\ref{tab:1}. In the present work only those parameters appearing in the fractional cell kill ($d$, $\lambda$ and $s$) are inspected. Accordingly to the CA model, we have set $\sigma=0$ in the ODE model since the CA does not include a constant input of effector cells. We have also selected a value $g=0.15$, which is very close to one of the values appearing in Table~\ref{tab:1}. Importantly the CA model and the ODE model include the same type of processes. The logistic growth of tumor cells in the CA model arises as a consequence of competition for nutrients \citep{ferreira}. There is also competition among healthy cells and tumor cells for nutrients, which in the ODE model is represented by the competing Lotka-Volterra terms between healthy and tumor cells. T cell lysis, inactivation and recruitment are also present in both models. Only the competition term between tumor and immune cells $a_{31}$ is different. Although we keep this parameter as shown in Table~\ref{tab:1}, if desired, it can be made equal to zero. As far as we have investigated, reducing the value of this parameter produces no appreciable consequences in our study. Notwithstanding this correspondence, we recall that during the second step of our CA simulations, the tumor dynamics is suspended. Accordingly, the parameter $r_{1}$ should be made equal to zero. Again, we keep this parameter as shown in Table~\ref{tab:1}. Reducing the value of this parameter produces no significant consequences in our study when the T cells are effective, because the time scale of T cell lysis (less than an hour) is considerably smaller that the time scale of cell-division (around a day). For immunodeficient scenarios the effects are more sensitive, but still small. In other words, T cell dynamics dominates during the first four hours. 
\begin{table}
\centering
\resizebox*{10.8cm}{10.0cm}{
\begin{tabular}{|c c l l|}
\hline
\rowcolor[gray]{0.95}[0.21cm][0.27cm]Parameter &Units  &Value &Description \\
\hline
$r_{1}$ &day$^{-1}$ &$5.14\times10^{-1}$ & \small Tumor cells growth rate \\
$K_{1}$ &cell & $9.8\times10^{8}$ & \small Tumor carrying capacity \\
$a_{12}$ &cell $^{-1}$ day $^{-1}$ &$1.1\times10^{-10}$ & \small Competition of host cells with tumor cells \\
$d(nn)$ &day$^{-1}$  & 2.20 & Saturation level of fractional tumor cell kill \\
$d(nl)$ & & 3.47 &  \\
$d(ln)$ & & 2.60 &  \\
$d(ll)$ & & 7.86 &  \\

  &   &   &   \\

$s(nn)$ &None & $1.6$ & Steepness coefficientof of fractional tumor cell kill \\
$s(nl)$ &  & $2.5$ & \\
$s(ln)$ &  & $1.4\times10^{-1}$ & \\
$s(ll)$ &  & $4.0\times10^{-1}$ & \\
$\lambda(nn)$ &None  & $1.2\times10^{-1}$ & Exponent of fractional tumor cell kill \\
$\lambda(nl)$ & & $2.1\times10^{-1}$ & \\
$\lambda(ln)$ & & $7.0\times10^{-1}$ & \\
$\lambda(ll)$ & & $7.0\times10^{-1}$ & \\

  &   &   &   \\

$r_{2}$ & day$^{-1}$ & $1.80\times10^{-1}$ & \small Host cells growth rate \\
$K_{2}$ & cell  & $1.0\times10^{9}$ & \small Host cells carrying capacity \\
$a_{21}$ &cell $^{-1}$ day $^{-1}$ &$4.8\times10^{-10}$ & \small Competition of tumor cells with host cells \\

 &  &  &  \\

$\sigma$ & cells day$^{-1}$ & $7.5\times10^{4}$ & \small Constant source of effector cells \\
$d_{3}$ & day$^{-1}$  & $6.12\times10^{-2}$ & \small Inactivation rate of effector cells \\
$g(nn)$ &day$^{-1}$ &$3.75\times10^{-2}$ & \small  Maximum recruitment rate \\
$g(nl)$ & & $3.75\times10^{-2}$ & \\
$g(ln)$ & & $1.13\times10^{-1}$ & \\
$g(ll)$ & & $3.00\times10^{-1}$ & \\
$h$ &cell$^{2}$ &$2.02\times10^{7}$ & \small Steepness coefficient for the recruitment\\
$a_{31}$ &cell $^{-1}$ day $^{-1}$ & $2.8\times10^{-9}$ & \small Immune-tumor competition \\
\hline
\end{tabular}}
\caption{The values of the parameters in the ordinary differential equation model used to fit experiments from \cite{nkg2d}. The parenthesis represent four different cases: a primary challenge with control-transduced cells followed by a secondary one with ligand (\emph{nl}) or control (\emph{nn}) cells, and a primary interaction with ligand-transduced cells followed again by ligand (\emph{ll}) or ligand-negative (\emph{ln}) rechallenges.}
\label{tab:1}
\end{table}

\section{Results}\label{sec:res}

\subsection{Tumors}

\begin{figure}
\centering
\includegraphics[width=0.8\linewidth,height=1.1\linewidth] {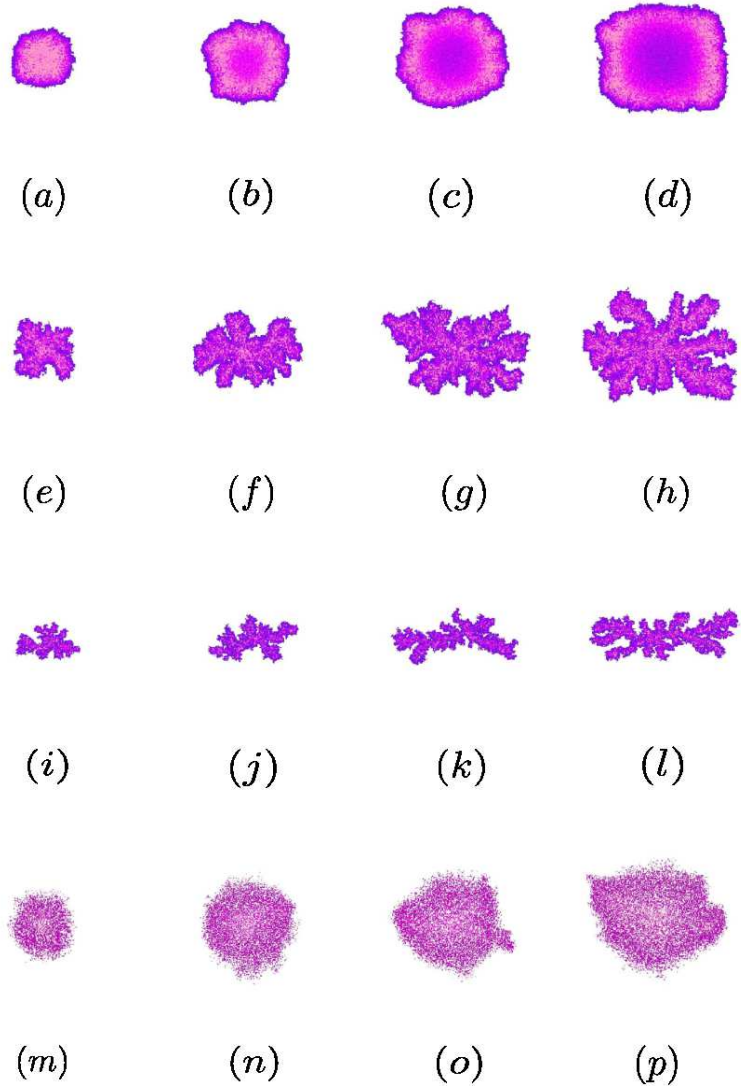}
\caption{Tumors generated using the cellular automaton model. Tumors become increasingly branchy as the competition for nutrients increases. Colors go from dark purple (one cell) to light pink (highest number of cells in a grid point for each tumor). We set the parameters $\lambda_{M}=10$ and $\theta_{nec}=0$ in all the cases, disregarding necrosis. (a-d) Spherical tumors with increasing size and parameters $\alpha=2/n$, $\lambda_{N}=25$, $\theta_{div}=0.3$ and $\theta_{mig}=\infty$. (e-h) Papillary tumors with increasing size and parameters $\alpha=4/n$, $\lambda_{N}=200$, $\theta_{div}=0.3$ and $\theta_{mig}=\infty$. (i-l) Filamentary tumors with increasing size and parameters $\alpha=8/n$, $\lambda_{N}=270$, $\theta_{div}=0.3$ and $\theta_{mig}=\infty$. (m-p) Disconnected tumors with increasing size and parameters $\alpha=3/n$, $\lambda_{N}=200$, $\theta_{div}=0.75$ and $\theta_{mig}=0.02$.}
\label{fig:tumors}
\end{figure}

In Fig.~\ref{fig:tumors} we depict the simulated solid tumors with four distinct morphologies, depending on the nutrient competition among tumor cells. The apparent three-dimensionality is an artefact resulting from the fact that we let cells pile up at the CA grid points. This piling mechanism was assumed in \cite{ferreira} for computational simplicity, and does not have any consequence in our study, since once the tumors are grown, we project them to study their lysis. High values of $\alpha$ and $\lambda_{N}$ lead to more branchy tumors, gradually changing from spherical to filamentary. This break of the spherical symmetry of the tumors is explained if we consider that when some nearby neoplastic cells on the boundary of a tumor compete aggressively for nutrients, those cells that divide and take ahead at some step, preserve this advantage at the next step, stealing the nutrients to those cells left behind. The four geometries are comparable to a variety of histologies \citep{ferreira}, such us a basal cell carcinoma, a scamous papyloma, a trichoblastoma and a plasmacytoma. Note that the necrosis of tumor cells due to the scarcity of nutrients in the core of the masses has been neglected, since it has no relevance in our study. In the CA this is achieved by setting $\theta_{nec}=0.01$ for all our simulations. Except for the disconnected patterns appearing in the last row in Fig.~\ref{fig:tumors}, motility has been also disregarded, considering sufficiently high values of $\theta_{mig}$.

\subsection{Effective immune response}

In the model given by Eqs.~\eqref{eq:3},~\eqref{eq:4} and~\eqref{eq:5}, the fractional cell kill of tumor cells by CTLs is given by the function $K(E,T)$. In \cite{validbul} we opted for expressing this function in the form
\begin{equation}
K(E,T)=d\frac{E^{\lambda}}{h(T) + E^{\lambda}},
\label{eq:7}
\end{equation}
with $h(T)=s T^{\lambda}$. Written this way, the fractional cell kill clearly states that the more effector cells, the greater the fractional cell kill, but bearing in mind the saturation of antigen-mediated immune response, which depends on the tumor burden. We propose that the saturation is due to the crowding of immune effector cells, which is evident if we recall that these cells need to be in contact with tumor cells to exterminate them. In a solid tumor, once all the tumor cells on its surface are in contact with a first line of immune cells, the remaining effector cells are not lysing, although the adjacent lines behind probably contribute to immune stimulation through several feedback mechanisms. Therefore, at a certain point, no matter how many more immune cells are present in the region of interest, the rate at which the tumor is lysed remains practically unaltered. Before saturation appears, if two tumors of the same nature and different size at a certain time instant are lysed at the same rate by the immune system, the bigger tumor will require more effector cells. Put more simply, if two tumors of different size are reduced to a particular fraction of its size after a certain period of time, the bigger tumor will require more effector cells. The number of effector cells $E$ for which the fractional tumor cell kill is half of its maximum $d$, increases monotonically with the tumor size $h(T)$.

\begin{figure}
\centering
\includegraphics[width=0.8\linewidth,height=1.1\linewidth] {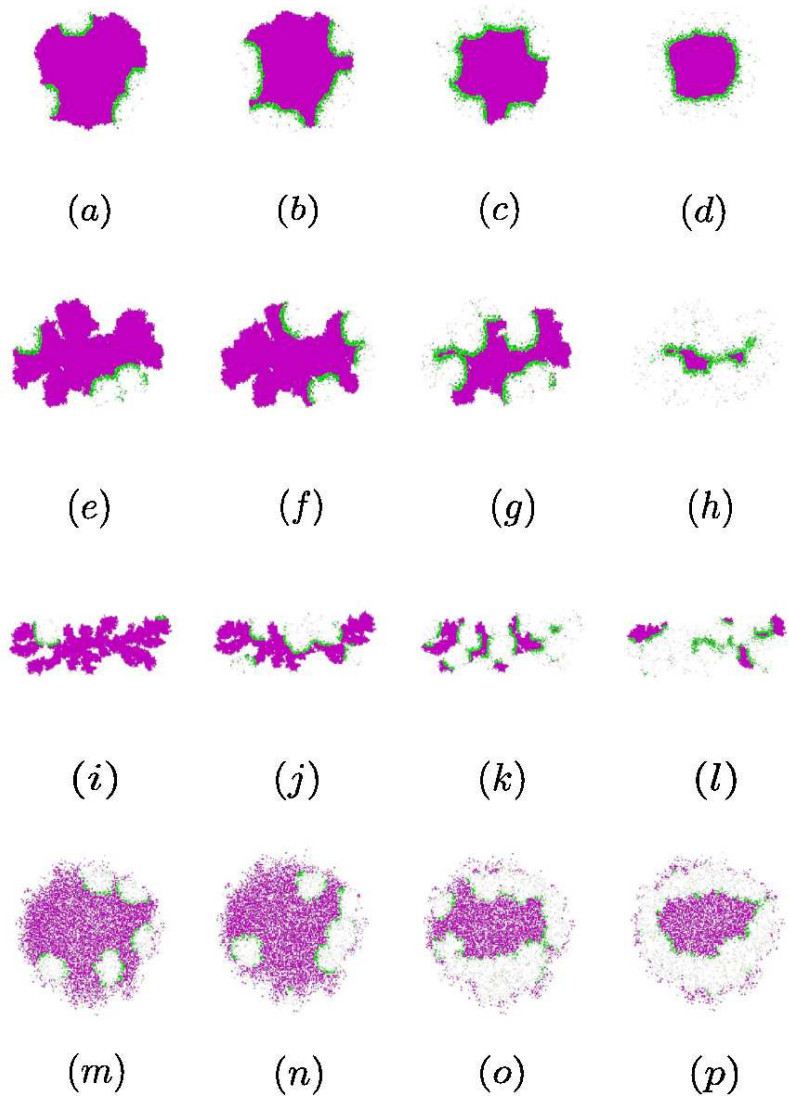}
\caption{Lysed tumors after four hours for different effector-to-target ratios. The effector cells (green) form satellites that advance destroying their neoplastic enemies (violet) and leave apoptotic bodies (light gray) behind them. The parameter values of the CA are $\theta_{lys}=0.3$, $\theta_{rec}=1.0$, $\theta_{inc}=0.5$, $\lambda_{M}=10$, $\lambda_{N}=25$ and $\alpha=2/L$. (a-d) Spherical tumor in Fig.~2(b) with $E_{0}/T_{0}$ taking values on the set $\{0.0025,0.005, 0.05, 0.75\}$, respectively. (e-h) Papillary tumor in Fig.~2(f) with $E_{0}/T_{0}$ taking values on the set $\{0.005, 0.0075, 0.025, 0.25 \}$, respectively. (i-l) Filamentary tumor in Fig.~2(l) with $E_{0}/T_{0}$ taking values on the set $\{0.0025,0.0075,0.017,0.025\}$, respectively. (m-p) Disconnected tumor in Fig.~2(n) with $E_{0}/T_{0}$ taking values on the set $\{0.005, 0.0075$, $0.015, 0.025\}$, respectively.}
\label{fig:lysed}
\end{figure}

We use simulations to demonstrate that these assertions are sufficient to explain the fractional cell kill law, even though there might be others. With this purpose, for every tumor pictured in the previous section, we prepare co-cultures with different effector-to-target ratios. Then, we let the CA evolve and measure the lysis four hours later (see Fig.~\ref{fig:lysed}). As previously explained, the tumors have been projected before the lysis starts, to better correlate the geometry and the parameters in the fractional cell kill. Otherwise, we would have two-dimensionally distributed lymphocytes fighting three-dimensional-like tumors, since in our CA we do not let the immune cells pile up. We do it this way to avoid the unfair situation in which just a few immune cells are facing a big pile of tumor cells, and vice versa. Finally, the results are fitted to the ODE model using a least-squares fitting method. We recall that such model was validated using as initial conditions typical cell populations of $10^{6}$ cells, while the CA automaton grid used can harbour at most $9\times10^{4}$ cells. However, this is not a hurdle at all, since if desired, the cell populations in the ODE model can be renormalized and its parameters redefined so as the cell numbers coincide.
\begin{table}
\centering
\resizebox*{10.9cm}{5.0cm}{
\begin{tabular}{|c c l l|}
\hline
\rowcolor[gray]{0.95}[0.21cm][0.21cm]Parameter &Units  &Value &Description\\
\hline
$d$(s) &day$^{-1}$  & $9 \pm 4$  & Saturation level of fractional tumor cell kill  \\
$d$(p) & & $20 \pm 1$ & \\
$d$(f) & & $32 \pm 2$ & \\
$d$(d) & & $13 \pm 3$ & \\
$\lambda$(s) &None  & $0.61 \pm 0.07$ & Exponent of of fractional tumor cell kill \\
$\lambda$(p) & & $0.87 \pm 0.04$ & \\
$\lambda$(f) & & $0.89 \pm 0.03 $ & \\
$\lambda$(d) & & $0.63 \pm 0.03$ & \\
$s$ &None & $0.15$ & Steepness coefficient of fractional tumor cell kill  \\

  &   &   &  \\

$D_{F}$(s) &None & $1.09 \pm 0.02$ & Box counting dimension of the boundary before the lysis starts \\
$D_{F}$(p) &  & $1.21 \pm 0.04$ & \\
$D_{F}$(f) &  & $1.36 \pm 0.02$ & \\
$D_{F}$(d) &  & $1.72 \pm 0.04$ & \\
\hline
\end{tabular}}
\caption{The parameter values modified in the model shown in Eqs.~\eqref{eq:3}, \eqref{eq:4} and \eqref{eq:5} corresponding to an effective immune response. The parameters $\lambda$ and $d$ are obtained through a least-square fitting of the lysis of tumor cells between the CA simulations and the ODE model. The mean value and standard deviations is computed for each morphology using four different tumors sizes: spherical (s), papillary (p), filamentary (f) and disconnected (d).}
\label{tab:2}
\end{table}

\begin{figure}
\centering
\includegraphics[width=0.8\linewidth,height=0.8\linewidth] {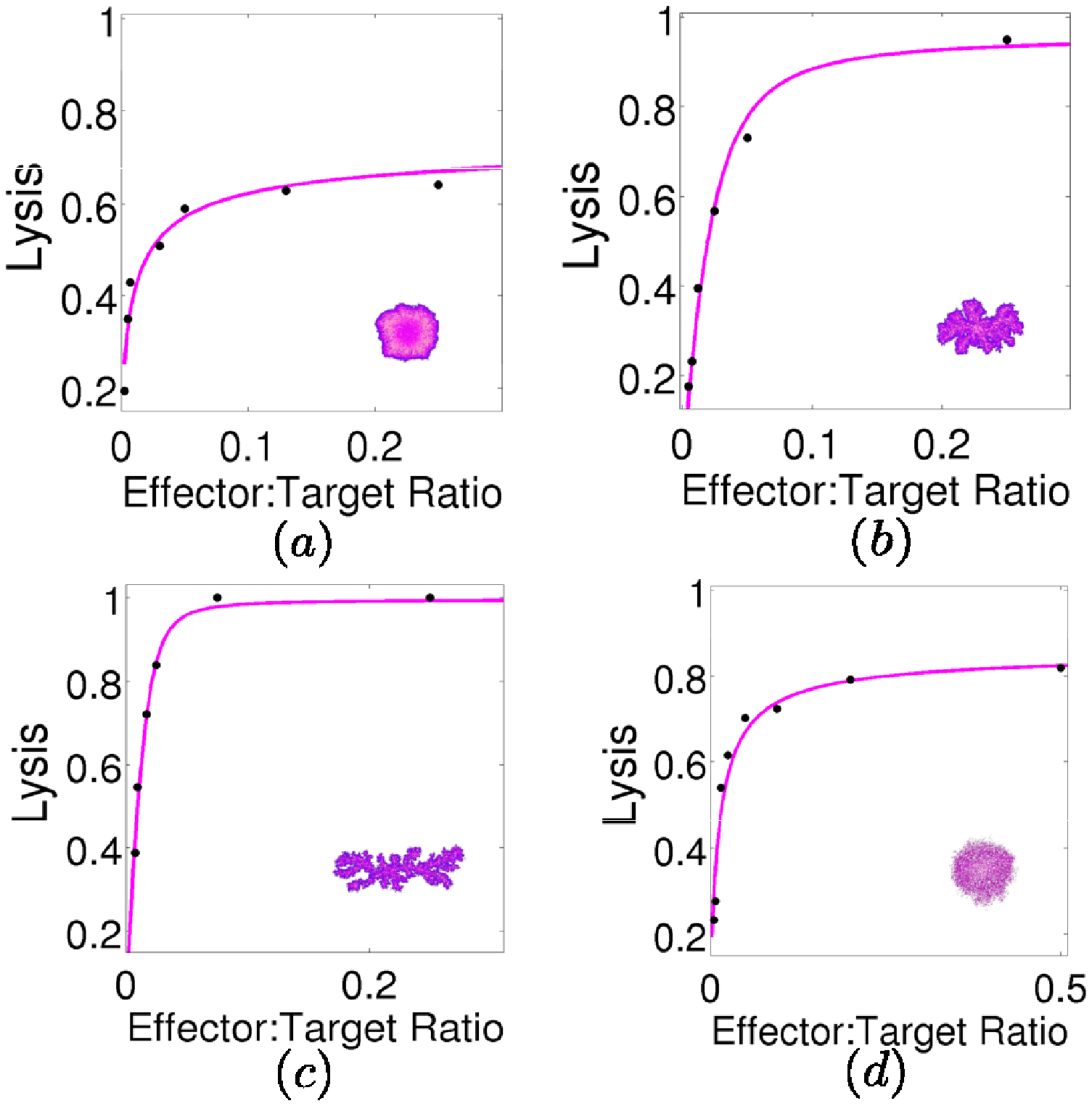}
\caption{The lysis of tumor cells after four hours versus the effector-to-target ratio $E_{0}/T_{0}$ in immunocompetent environments. The parameter values of the CA related to the lysis, recruitment and inactivation are $\theta_{lys}=0.3$, $\theta_{rec}=1.0$ and $\theta_{inc}=0.5$, respectively. The solid curve corresponds to the ODE model, while the points correspond to the cellular automaton results. (a) The spherical tumor in Fig.~2(b). (b) The papillary tumor in Fig.~2(f). (c) The filamentary tumor in Fig.~2(l). (d) The disconnected tumor in Fig.~2(n).}
\label{fig:ajustelig}
\end{figure}

The resulting lysis curves are depicted in Fig.~\ref{fig:ajustelig} and the values of the parameters $d$, $\lambda$ and $s$ in Eq.~\eqref{eq:7} are listed in Table~\ref{tab:2}, together with the fractal dimension $D_{F}$ of the boundary of the initial tumors. Satellitosis is clearly appreciated as a consequence of T cell recruitment, and the resulting clusters of cells act like wave fronts that advance lysing the tumor. Note that the immune cells that are far enough from the tumor become inactivated after several iterations of the CA. Consequently, only the T cells that are able to make contact with the tumor, gain traction in killing and subsequent recruitment, appear in the figures. There is a correlation between the box counting dimension and the parameters $d$ and $\lambda$ for the connected tumors examined, but this is not case for the disconnected one. The disconnected tumors shown in Fig.~\ref{fig:tumors} display the highest box-counting dimension, because they are very drilled, so that most of the tumor cells are on its boundary. However, they are rather spherical, and for this reason the part of the boundary that is in the center of the mass is not initially accessible to the immune cells. These facts explain the low values of $d$ and $\lambda$ for such tumors, which are comparable to the spherical ones. Therefore, in our model, those tumors with a bigger surface of contact are lysed faster. Indeed, what matters to the cytotoxic cells is how accessible their enemies are. The more tumor cells there are between an immune cell and some other tumor cell, the lower the rate at which the effector cells kill their victims. This is starkly evident for the spherical tumors, which correspond to the smallest values of $d$ and $\lambda$. 

Thus, according to our model, Eq.~\eqref{eq:7} is a robust emergent property of the tumor-immune interaction depending on the spatial distribution of the tumor cells. It reflects the tumor size dependent saturation of an effective immune system, fruit of the crowding of the effector cells and the arduousness to establish contact with their adversaries. Nevertheless, it takes hours for the effector cells to fully lyse the tumors so far investigated, what denotes that this extrinsic limitation to the lytic capacity of the immune system is barely important compared to the immunoevasive manoeuvres that tumor cells commonly orchestrate \citep{hallmarks}.

\subsection{Ineffective immune response}

Tumor cells find ways to evade the immune surveillance through a broad range of mechanisms \citep{biocan}. They can acquire the ability to repress tumor antigens, MHC class I proteins or NKG2D ligands. They may also learn to destroy receptors or to saturate them, induce suppressor T cells formation, launch counterattacks against immunocytes by releasing cytokines, avoid apoptosis, etc. It is therefore pertinent to ask ourselves if the fractional cell kill can cover situations in which the tumor microenvironment is immunodeficient. 

In \cite{validpilis} the authors show that the lysis curves corresponding to NK cells in the experiments borrowed from \cite{nkg2d} do not show saturation, and that a fractional cell kill given by a simple power law $cE^{\nu}$ works to fit such data. Because much higher values of the effector-to-target ratio are required to obtain similar values for the lysis compared to the CTLs curves, it was suggested that when the effector cells are less effective, saturation is not observed. 

\begin{figure}
\centering
\includegraphics[width=1.0\linewidth,height=0.5\linewidth] {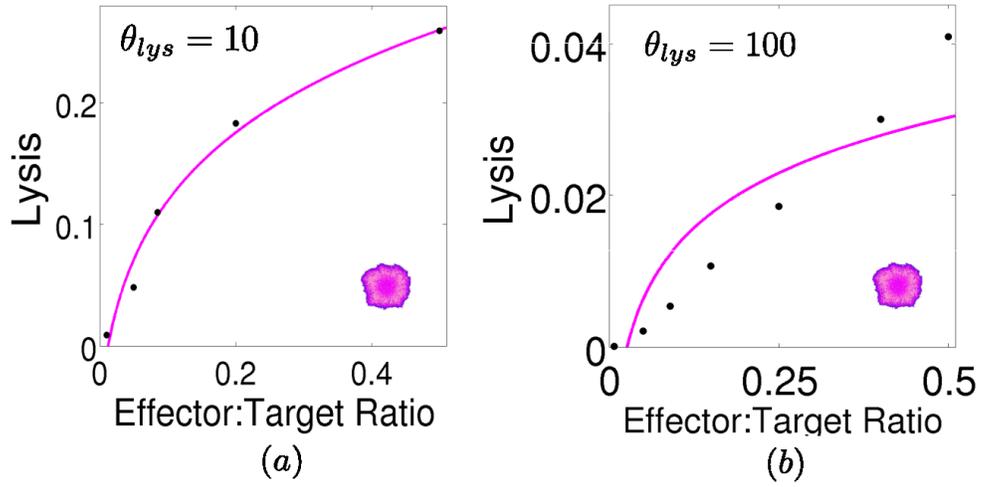}
\caption{The lysis of tumor cells after four hours versus the effector-to-target ratio $E_{0}/T_{0}$ in immunosuppressed environments. The spherical tumor represented in Fig.~2(b) is studied, with recruitment and inactivation CA parameters $\theta_{rec}=1.0$ and $\theta_{inc}=0.5$. The solid curve corresponds to the ODE model, while the points correspond to the cellular automaton results. (a) A more ineffective, but still effective, adaptive response is here represented, with $\theta_{lys}=10$. (b) A value of the intrinsic cytotoxic capacity $\theta_{lys}=100$ is set for the most ineffective immune system.}
\label{fig:lysnoneffec}
\end{figure}

Mathematical arguments have been given \citep{validbul} to explain this lack of saturation. Briefly, when the cytotoxic cells are less effective, only a fraction $f$ of the effector cells are interacting with the tumor. Thus we can replace $E$ by $f E$ in the fractional cell kill. Now, defining $\tilde{s}=s /f^{\lambda}$, the fractional cell kill law remains unchanged. This suggests that the parameter $s$ is related to the effectiveness of the cytotoxic cells, being this parameter inversely proportional to the effectiveness of such cells. On the other hand, if the effectiveness is small enough ($f\ll1$), then $h(T)$ dominates over $E^{\lambda}$ in Eq.~\eqref{eq:7}, as long as $E$ is not too high. The resulting lysis term becomes $d f^{\lambda} E^{\lambda} T^{1-\lambda}/s$. This facts legitimize the estimation $c E^{\nu}T$ that has been used in other works \citep{validpilis,validbul} to reproduce the fractional cell kill of tumor cells. Nevertheless, here we do not want to introduce phenomenological functions of this type, but rather concentrate our efforts on the significance of $s$. To this end, we diminish the intrinsic cytotoxic capacity of the immune cells, which is encoded in the parameter $\theta_{lys}$ in our cellular automaton. Higher values of this parameter represent more ineffective T cells. The results can be seen in Fig.~\ref{fig:lysnoneffec} and the values of the parameters are listed in Table~\ref{tab:3}. As we increase the parameter $\theta_{lys}$, the saturation appearing in the lysis curves become less evident, and at a certain point it disappears. 

When $\theta_{lys}=10$, the ODE model can be adjusted to the CA results. However, increasing $s$ is not sufficient to reproduce this data, and considerable variations of the remaining parameters $d$ and $\lambda$ is required. A much more dramatic case arises when $\theta_{lys}=100$. In this case we have not been able to find any values of the parameters that represents faithfully the CA results. The best fitting provided by the ODE model exhibits considerable saturation. The conclusion is that the fractional cell kill represented by Eq.~\eqref{eq:7} works bad for immunodefficient environments and also confuses the geometrical effects and the intrinsic cytotoxic capacity of the immune cells. In the next section, we propose a new fractional cell kill that allows to fit the results more accurately by simply reducing the value of $s$.

\subsection{Modification of the fractional cell kill}

In \cite{validbul}, the particular nature of the function $h(T)$ appearing in Eq.~\eqref{eq:7} was also discussed, proving that if instead of $h(T)=s T^{\lambda}$, $h(T)=s T^{\lambda+\Delta \lambda}$ is used, the empirical results can also be validated by simply decreasing the value of $s$, even for values $\Delta \lambda/\lambda$ greater that one. This means that the original proposal of a saturating fractional cell kill depending on the quotient $E/T$ can not be guaranteed.

\begin{table}
\centering
\resizebox*{11.9cm}{3.0cm}{
\begin{tabular}{|c c l l|}
\hline
\rowcolor[gray]{0.95}[0.17cm][0.3cm]Parameter &Units  &Value &Description\\
\hline
$d$(s) &day$^{-1}$ & $3.80$ & Saturation level of the fractional tumor cell kill \\
$d$(i) & & $1.56$ & \\
$\lambda$(s) &None  & $0.62$ & Exponent of the fractional tumor cell kill \\
$\lambda$(i) &  & $0.17$ &  \\
$s$(s) &None & $0.50$ & Steepness coefficient of the fractional tumor cell kill \\
$s$(i) & & $1.10$ &  \\

\hline
\end{tabular}}
\caption{The parameter values of the fractional cell kill given by Eq.~\eqref{eq:7}. These parameters are obtained through a least-square fitting of the lysis of tumor cells between the CA simulations and the ODE model (see Fig.~\ref{fig:lysnoneffec}. Two cases are represented: a very ineffective (i) and a semi-effective (s) immune responses.
}
\label{tab:3}
\end{table}

Furthermore, from a theoretical point of view, the function $h(T)=sT^{\lambda}$ makes the model ill-defined in the limit of very big tumors ($T \rightarrow \infty$) facing a comparably small fixed number of immune cells. The reason is that in this limit we get unbounded velocity for the lysis ($K(E,T)T \rightarrow  \infty$). We demonstrate that $h(T)=s T$ is a much better choice. It has been shown \citep{tils, micaelis} that for a fixed number of effector cells $E_{0}$, the Michaelis-Menten kinetics govern the lysis of tumor cells. The  value of the lytic velocity at tumor saturation, \emph{i.e.}, when $T \rightarrow \infty$, is reported in such works as a measure of the intrinsic cytotoxic capability of a particular number of effector cells. A Michaelis-Menten decay in Eq.~\eqref{eq:3} is obtained for a constant value of effector cells as long as $h(T)=sT$ in used. The value at saturation for a fixed number of effector cells is then $d E_{0}^{\lambda}/s$. An argument supporting saturation comes from the following fact. If the number of tumor cells is much higher than a fixed number of effector cells, the velocity at which the tumor cells are lysed can not be enhanced by increasing the number of the neoplastic cells. This occurs because T cells kill tumor cells one by one, and for such ratios all the effector cells are already busy fighting other cells. In a similar fashion, for an enzymatic reaction, one can not increase arbitrarily the velocity at which the products are formed by simply adding more substrate. Precisely, this reasoning is reminiscent of the original formulation proposed by \cite{immutu}, in which the cell populations are regarded as chemical species obeying enzymatic kinetics in the quasi-steady state regime. In such work, the tumor cells are the substrate, the effector cells are the enzyme and the products are the dead cells. Indeed, in Appendix \ref{app:foobar} we use enzyme kinetics as a metalanguage to provide an analytical derivation of the fractional cell kill. A fractional cell kill function that yields bounded velocity for the lysis of tumor cells when any of these two cell populations is sufficiently high compared to the other is represented by
\begin{equation}
K(E,T)=d\frac{E^{\lambda}}{s T + E^{\lambda}}.
\label{eq:8}
\end{equation}

If we focus only on the lysis of tumor cells, the velocity at which the tumor is reduced can be represented by the following nonlinear differential equation
\begin{equation}
\dot{T}=-d \dfrac{E^{\lambda}}{s T+E^{\lambda}}T. 
\label{eq:2}
\end{equation}

\begin{figure}
\centering
\includegraphics[width=1.0\linewidth,height=1.0\linewidth] {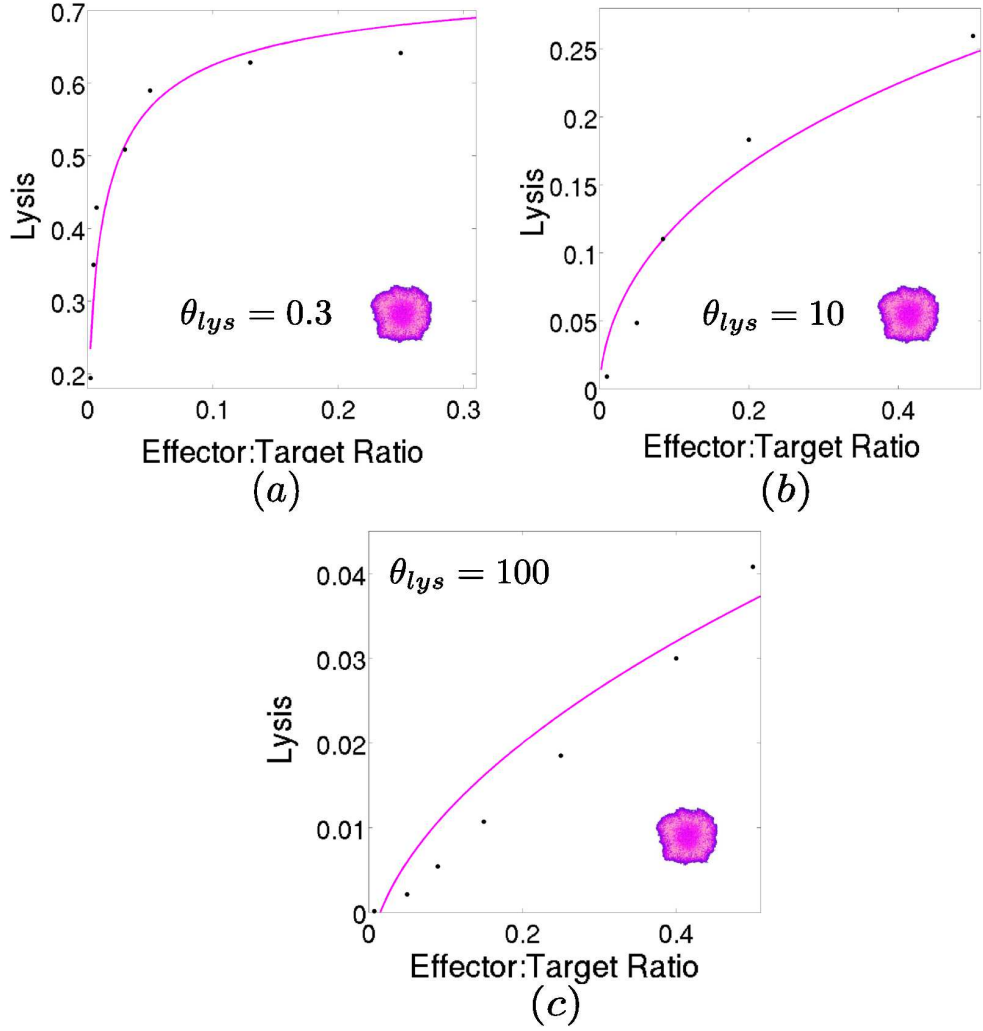}
\caption{The lysis of tumor cells after four hours versus the effector-to-target ratio $E_{0}/T_{0}$ for increasing ineffectiveness of the lymphocytes. The spherical tumor represented in Fig.~2(b) is studied, with recruitment and inactivation CA parameters $\theta_{rec}=1.0$ and $\theta_{inc}=0.5$. The solid curve corresponds to the ODE model, while the points correspond to the cellular automaton results. (a) An effective immune response for $\theta_{lys}=0.3$. (b) A more ineffective, but still effective, adaptive response is here represented, with $\theta_{lys}=10$. (c) A value of the intrinsic cytotoxic capacity $\theta_{lys}=100$ is set for the most ineffective immune system. As shown in Tab.~\ref{tab:4}, the intrinsic cytotoxic potential of the T cells is chiefly represented by parameter $s$ in Eq.~\eqref{eq:8}.}
\label{fig:6}
\end{figure}

Following the point of view of \cite{immutu}, this mathematical expression can be regarded as a Michaelis-Menten kinetics where the rate constants of the formation of the ``enzyme-substrate" conjugates, their dissociation and their conversion to product depend nonlinearly (as power laws) on the enzyme concentration. It establishes the saturation of the velocity of the lysis of tumor cells for both the tumor and the immune cell populations. In Fig.~\ref{fig:6}(a) we first reproduce the experiments of the spherical tumor shown in Fig.~2(b) for $\theta_{lys}=0.3$. This allows us to obtain the parameter values of the modified fractional cell kill shown in Eq.~\eqref{eq:8}. Then we carry out the simulations of the preceding section for immunodeficient environments and see how, mainly by increasing the value of $s$, the CA results are reproduced (see Figs.~\ref{fig:6}(b) and ~\ref{fig:6}(c)). The parameter values are listed in Table~\ref{tab:4}. This sheds light into the significance of this parameter, which is now manifestly related to the intrinsic cytotoxic potential of the T cells. Moreover, this implies that the limit $T \rightarrow \infty$, for which the quantity $d E^{\lambda}/s$ is obtained, is not a good measure of lymphocyte cytotoxicity, as suggested in \cite{tils, micaelis}. This limit, which for a constant value of the T cells implies a linear decay of the tumor, involves geometry as well. Ideally, if we consider that there is just one immune cell, and it takes this cell an hour to lyse a tumor cell, then a spherical tumor would be reduced at approximately one cell per hour (assuming that this immune cell does not become inactivated at some step). However, the geometry of the tumor, which is coded in the parameters $d$ and $\lambda$, clearly affects how fast this single cell can erase it. 

Even though the reduction of saturation for ordinary values of the effector-to-target ratio can be justified mathematically and numerically, the change in curvature for the CA results appearing in Fig.~\ref{fig:6}(c) requires a positive feedback mechanism. Certainly, the mechanism responsible for this phenomenon is the recruitment of immune cells, which becomes increasingly important as the effectiveness of the T cells decreases.

\begin{table}
\centering
\resizebox*{11.2cm}{4.0cm}{
\begin{tabular}{|c c l l|}
\hline
\rowcolor[gray]{0.95}[0.17cm][0.3cm]Parameter &Units  &Value &Description\\
\hline
$d$(e) &day$^{-1}$ & $9.22$ & Saturation level of the fractional tumor cell kill \\
$d$(s) & & $9.62$ &  \\
$d$(i) & & $9.52$ & \\
$\lambda$(e) &None  & $0.50$ & Exponent of the fractional tumor cell kill \\
$\lambda$(s) &  & $0.51$ &  \\
$\lambda$(i) &  & $0.55$ &  \\
$s$(e) &cells$^{\lambda-1}$ & $1.0\times10^{-5}$ & Steepness coefficient of the fractional tumor cell kill \\
$s$(s) & & $1.4\times10^{-4}$ &  \\
$s$(i) & & $9.5\times10^{-4}$ &  \\

\hline
\end{tabular}}
\caption{The parameter values of the fractional cell kill appearing in Eq.~\eqref{eq:8}. These parameters are obtained through a least-square fitting of the lysis of tumor cells between the CA simulations and the ODE model (see Fig.~\ref{fig:6}). Three cases are represented: an effective (e), a semi-effective (s) and ineffective immune responses (i). Note that it is only the parameter $s$, which is related to the intrinsic cytotoxic capacity, that varies substantially. It increases as the immune cells become less effective.}
\label{tab:4}
\end{table}

\section{Discussion} \label{sec:dis}

Our study demonstrates that the saturation of the fractional cell kill of tumor cells by their cytotoxic opponents is a consequence of cell crowding. This limitation depends on the morphology of the tumor, insofar as geometry restricts the access of effector cells to tumor cells. In theory, those tumor growing with ``spherical symmetry" will be the harder to lyse, because many layers of tumor cells have to be erased to reach the cells at the center. We recall that the process of T cell recruitment from circulation to the tumor site is complex, involving several steps \citep{recr}. This implies that the crowding might happen before contact with the tumor occurs, as for example, during adhesion to the endothelium. In such a case, a relation between the parameters in the fractional cell kill and the shape of the tumor can not be established. At all events, mathematically, this extrinsic barrier to the lytic capacity of the effector cells is reflected in the parameters $d$ and $\lambda$. Less spherical tumors correspond to higher values of both parameters. Interestingly, the values of $\lambda$ are expected to be between zero and one, as suggested by the experiments and the simulations. From the enzymatic kinetics point of view, if we think of Eq.~\eqref{eq:8} as a Hill function depending on the effector cells, this can be interpreted as non-cooperative binding. Certainly, if we pay attention to the process of lysis only, the best that an immune cell can do to another is not to interpose between itself and their adversaries. Of course, cooperative effects exist, as the recruitment term exemplifies. Quite the opposite, as the intrinsic lytic capacity of cytotoxic cells is decreased, saturation gradually vanishes. This capability is inversely proportional to the parameter $s$. It is not surprising that the limit $d E^{\lambda}/s$ dictates the lysis when the immune response is ineffective. This occurs because saying that a small fraction of a cytotoxic cell population $f E$ interacts with a tumor is equivalent to considering a small number of effector cells confronting a big tumor.

\section*{Aknowledgments}  \label{sec:aknow}

This work has received financial support by the Spanish Ministry of Economy and Competitivity under project number FIS2013-40653-P. 

\begin{appendices}
 \renewcommand{\theequation}{A\arabic{equation}}
 \setcounter{equation}{0} 
\section{Cellular automaton rules} \label{app:caraa}
  
The CA rules are described for the two steps, one corresponding to the development of the tumors, and the other related to the lysis of the tumor cells by the cytotoxic T cells. They are almost the same as those used in \cite{cahyb}, and any difference will be explicitly remarked. In what follows, $T(\vec{x})$ and $E(\vec{x})$ are the tumor and immune cells at position $\vec{x}$, while $N(\vec{x})$ and $M(\vec{x})$ are the concentration of nutrients in nondimensional variables at position $\vec{x}$. $N(\vec{x})$ represents those nutrients required for cell division, and $M(\vec{x})$ those required for other cellular activities. The parameters $\theta_{a}$ represent the intrinsic capacity of a cell to carry out a particular action $a$.
  
\subsection{First step} 
As in previous works, the role of the healthy cells is simplified to passive competitors for nutrients that allow the tumor cells to freely divide or migrate. At each CA iteration the tumor cells are randomly selected one by one, and a dice is rolled to choose weather each of these cell divides (1), migrates (2) or dies (3).
\begin{enumerate}
\item A tumor cell divides with probability
\begin{equation}
P_{div}=1-\exp\left(-\dfrac{(N/T)^{2}}{\theta_{div}^{2}}\right).
\end{equation}
    
This probability is compared to the probability of a randomly generated number using a normal distribution and the same standard deviation. If the former is greater than the last, division takes place. The higher the value of $\theta_{div}$, the more metabolic requirements for a cell to proliferate. When a cell at position $\vec{x}=(x,y)$ divides, if there are neighbouring CA elements that are not currently occupied by tumor cells, we randomly select one $\vec{x'}=(x',y')$ and place there the newborn cell, thus making $T(\vec{x'})=1$ and $H(\vec{x'})=0$ or $D(\vec{x'})=0$, where $D(\vec{x})$ is the function representing the necrotic cells at position $\vec{x}$. However, if all the neighbouring elements are occupied, we let the cells pile up, making $T(\vec{x}) \rightarrow T(\vec{x})+1$. 
\item A tumor cell migrates with probability
\begin{equation}
P_{mig}=1-\exp\left(-\dfrac{(\sqrt{T}M)^{2}}{\theta_{mig}^{2}}\right).
\end{equation}
    
If $P_{mig}$ is greater than the probability of a randomly generated number, migration proceeds, otherwise it does not. The higher the value of $\theta_{mig}$, the more metabolic requirements for a cell to migrate, unless there are too many tumor cells. When a cell at position $\vec{x}$ moves, if there are neighbouring CA elements that are not currently occupied by tumor cells, we randomly select one at $\vec{x'}$ and place the cell there. If there is more than one cell in the original position, the moving cell simply replaces the healthy or the necrotic cell, thus making the transformation $T(\vec{x}) \rightarrow T(\vec{x})-1$, $T(\vec{x'})=1$ and $H(\vec{x'})=0$ or $D(\vec{x'})=0$. 
\item On the other hand, if there is only one tumor cell at $\vec{x}$, then it interchanges its position with the healthy or necrotic cell at $\vec{x'}$. If all the neighbouring elements are occupied, we displace the cell to a randomly selected neighbouring element.
\item A tumor cell dies with probability
\begin{equation}
P_{nec}=\exp\left(-\dfrac{(M/T)^{2}}{\theta_{nec}^{2}}\right).
\end{equation}
        
If $P_{nec}$ is higher than the probability of a randomly generated number, necrosis proceeds, otherwise it does not. The higher the value of $\theta_{nec}$, the greater the probability for a cell to die. When a cell at position $\vec{x}$ dies, we make $T(\vec{x}) \rightarrow T(\vec{x})-1$. If this is the only cell at $\vec{x}$, then $D(\vec{x})=1$.
\end{enumerate} 
    
\subsection{Second step}

At each CA iteration the immune cells that have one or more tumor cells as first neighbors, carry out an attempt to lyse a randomly chosen surrounding tumor cell. This process occurs with probability
\begin{equation}
P_{lys}=1-\exp\left(-\dfrac{1}{\theta_{lys}^{2}}\left(\sum_{i\in\eta_{1}}E_{i}\right)^{2}\right),
\end{equation}
where $\eta_{n}$ indicates summation up to the $n$-th nearest neighbours. If $P_{lys}$ is higher than the probability of a randomly generated number, then the selected tumor cell dies. Therefore $T(\vec{x'})=0$, $D(\vec{x'})=1$ and the immune cell counter decreases by a unit.  
If the counter reaches a value of zero, it dies and it is replaced by a healthy cell. The smaller the value of $\theta_{lys}$, the greater the probability for an effector cell to lyse a tumor cell. This parameter was not present in \cite{cahyb} and is introduced here to model the intrinsic cytotoxicity of T cells. When a tumor cell is destroyed by an immune cell, the first neighbouring cells are flagged for recruitment. For each CA element without tumor cells a new immune cell is born with probability
\begin{equation}
P_{rec}=\exp\left(-\dfrac{1}{\theta_{rec}^{2}}\left(\sum_{i\in\eta_{1}}T_{i}\right)^{-2}\right).
\end{equation}

If $P_{rec}$ is higher than the probability of a randomly generated number, recruitment proceeds. The higher the value of $\theta_{rec}$, the less surrounding tumor cells that are required for T cell recruitment to success. When a cell is recruited at position $\vec{x'}$, we make $D(\vec{x'})=0$ or $H(\vec{x'})=0$, and $E(\vec{x'})=1$. 
        
Those effector cells which immediate neighbourhood is not occupied by tumor cells, either migrate or become inactivated. To decide which of these two processes is carried out, a coin is flipped. If the output is migration, it occurs for sure. In the opposite case, inactivation occurs with probability
\begin{equation}
P_{inc}=1-\exp\left(-\dfrac{1}{\theta_{inc}^{2}}\left(\sum_{i\in\eta_{3}}T_{i}\right)^{-2}\right).
\end{equation}
If $P_{inc}$ is higher than the probability of a randomly generated number, inactivation proceeds. The smaller the value of $\theta_{inc}$, the less surrounding tumor cells that are required for a T cell to become inactivated. When a cell disappears from position $\vec{x}$, we simply make $H(\vec{x})=1$ and $E(\vec{x})=0$.
    
 \renewcommand{\theequation}{B\arabic{equation}}
 \setcounter{equation}{0} 
\section*{APPENDIX B. The fractional cell kill as a Michaelis-Menten kinetics} \label{app:foobar}

The fractional cell kill represented by Eq.~\eqref{eq:8} can be derived from the Michaelis-Menten kinetics \citep{micamen,mimen} assuming that the rate constants of the reaction depend on the enzyme concentration. During the process of lysis, the effector cells $E$ bound to the tumor cells $T$ forming complexes $C$, and dead tumor cells $T^{*}$ result from this interaction. Therefore, the tumor cells play the role of the substrate and the effector cells act as the enzyme. This cellular reaction can be written in the form
\begin{equation}
E+T \underset{k_{-1}}{\overset{k_{1}}{\rightleftharpoons}} C \underset{k_{-2}}{\overset{k_{2}}{\rightleftharpoons}} T^{*}+E.
\label{eq:b1}
\end{equation}

Once a tumor cell is induced to apoptosis it can not resurrect, so we must set $k_{-2}=0$. Generally, also the backward reaction represented by $k_{-1}$ should be disregarded, since after tumor cell recognition and complex formation, destruction proceeds. However, we keep this term for reasons explained bellow. 

Assuming that the law of mass action holds, the system of differential equations governing the reactions is
\begin{eqnarray}
&\dfrac{d [E]}{d t}&=-k_{1}[E][T]+(k_{-1}+k_{2})[C] \label{eq:b2}\bigskip\\
&\dfrac{d [T]}{d t}&=-k_{1}[E][T]+k_{-1}[C] \label{eq:b3}\bigskip\\
&\dfrac{d [C]}{d t}&=k_{1}[E][T]-(k_{-1}+k_{2})[C] \label{eq:b4}\bigskip\\
&\dfrac{d [T^{*}]}{d t}&=k_{2}[C]. \label{eq:b5}
\end{eqnarray}

The Briggs-Haldane \citep{brha} quasi-steady state approximation $\dot{[C]}=0$ was assumed in \cite{immutu}. This approximation requires
\begin{equation}
 \dfrac{[E_{0}]}{[T_{0}]+K_{M}}\ll1,
\label{eq:b6}
\end{equation}
where $K_{M}=(k_{-1}+k_{2})/k_{1}$ is the Michaelis constant, and $[E_{0}]$ and $[T_{0}]$ are the initial concentrations of the effector and the tumor cells respectively.

Because we are dealing with situations in which the substrate concentration can be smaller than the enzyme, the quasi-steady state approximation implies $K_{M}\gg[E_{0}]$. Since this condition can not be generally guaranteed, instead, we consider Michaelis and Menten original formulation, and suppose that the substrate is in instantaneous equilibrium with the complex. We believe this is more reasonable, because it takes about an hour for a cytotoxic T cell to fully lyse a tumor cell and, if the cells are effective, the recognition and complex formation should occur quite fast when brought together. In this manner, we have $k_{1}[E][T]=k_{-1}[C]$. From Eqs.~\eqref{eq:b2} and~\eqref{eq:b4} we get the conservation law $[E]+[C]=[E_{0}]$. These two equations put together and substituted in Eq.~\eqref{eq:b5} yield
\begin{equation}
\dfrac{d [T^{*}]}{d t}=k_{2}k_{1}[E_{0}]\dfrac{[T]}{k_{1}[T]+k_{-1}}.
\label{eq:b7}
\end{equation}

So far, this is nothing else but the Michaelis-Menten kinetics. It is at this point that we have to consider a dependence of the rate constants of the reaction on the concentration of the effector cells. The mathematical relations are derived heuristically, based on the idea that for higher concentrations of the immune cells the rate constants vary in a such a manner that the reaction is pushed backwards. Since saturation is due to crowding of T cells, and this depends on the geometry of the tumor, it seems a natural choice to use power laws.

Once the first lines of effector cells cover the surface of a solid tumor, the remaining immune cells are not in contact with it. Alternatively, an equivalent argument is attained if we suppose that the non-interacting effector cells do interact with some tumor cells unsuccessfully (say ghost tumor cells), so that the complexes are dissociated without lysis. The more effector cells, the higher the rate of dissociation, and when the number of effector cells is small compared to the number of tumor cells, the dissociation should vanish. Therefore, we consider a power law dependence $k_{-1}([E_{0}])=\kappa_{-1}[E_{0}]^{\alpha}$, with $0<\alpha<1$, as suggested from the experiments. Substitution in Eq.~\eqref{eq:b7} yields
\begin{equation}
\dfrac{d [T^{*}]}{d t}=k_{2}k_{1}\dfrac{[E_{0}]}{k_{1}[T]+\kappa_{-1}[E_{0}]^{\alpha}}[T].
\label{eq:b8}
\end{equation}

The fractional cell production of dead cells in this equation already resembles very much to Eq.~\eqref{eq:8}. To obtain the exact result we have to consider dependence of $k_{1}$ and $k_{2}$ on the effector concentration as well. Note that for the inverse reaction to take place complexes have to be formed first, and this requires some time. Therefore, saying that complexes dissociate without lysis is not exactly equivalent to stating that the complexes are not formed. These rates should decay for increasing concentrations of the effector cells, diminishing the rate of formation of complexes and products. Once again, we postulate power law relations in the form $k_{1}([E_{0}])=\kappa_{1}[E_{0}]^{-\beta}$ and $k_{2}([E_{0}])=\kappa_{2}[E_{0}]^{-\gamma}$ where again $0<\beta<1$ and $0<\gamma<1$. It might result surprising that in the limit $[E_{0}]\rightarrow\infty$ these functional relations tend to zero, suggesting that the reaction stops. However, this is not the case, because when substituted in Eqs.~\eqref{eq:b2},~\eqref{eq:b3},~\eqref{eq:b4} and~\eqref{eq:b5}, $k_{1}([E])[E]$ and $k_{2}([E])[C]$ both increase with the number of effector cells. Replacing the rate functions in Eq.~\eqref{eq:b8} we obtain
\begin{equation}
\dfrac{d [T^{*}]}{d t}=\dfrac{\kappa_{1}\kappa_{2}}{\kappa_{-1}}\dfrac{[E_{0}]^{1-\gamma}}{\dfrac{\kappa_{1}}{\kappa_{-1}}[T]+[E_{0}]^{\alpha+\beta}}[T].
\label{eq:b9}
\end{equation}

We now rename the constants $\lambda=\alpha+\beta$,  $s=\kappa_{1}/\kappa_{-1}$, $d=\kappa_{1}\kappa_{2}/\kappa_{-1}$, and remember that the velocity for the lysis must remain bounded for $[E_{0}]\rightarrow\infty$, what imposes the constraint $\alpha+\beta+\gamma=1$. Thus, the velocity at which dead tumor cells accumulate is given by the nonlinear function
\begin{equation}
\dfrac{d [T^{*}]}{d t}=d \dfrac{[E_{0}]^{\lambda}}{s [T]+[E_{0}]^{\lambda}}[T].
\label{eq:b10}
\end{equation}

\end{appendices}

\end{document}